\documentclass[reprint,12pt]{elsarticle}

\usepackage{amssymb}
\usepackage{graphicx}
\usepackage{gensymb}
\usepackage{multirow}
\usepackage{multicol}
\usepackage{booktabs}
\usepackage{url}

\usepackage{breakurl}
\usepackage[breaklinks]{hyperref}

\journal{Carbon}

\makeatletter
\def\ps@pprintTitle{%
  \let\@oddhead\@empty
  \let\@evenhead\@empty
  \let\@oddfoot\@empty
  \let\@evenfoot\@oddfoot
}
\makeatother

\begin{document}
\begin{frontmatter}


\title{Aromatic molecules as sustainable lubricants explored by ab initio simulations}

\author[inst1]{Stefan Peeters}
\affiliation[inst1]{organization={Department of Physics and Astronomy, University of Bologna},
            addressline={Viale Carlo Berti Pichat 6/2}, 
            city={Bologna},
            postcode={40127}, 
            country={Italy}}
            
\author[inst1]{Gabriele Losi}
            
\author[inst2]{Sophie Loehl\'e}
\affiliation[inst2]{organization={TotalEnergies, OneTech},
            addressline={Chemin du Canal BP 22}, 
            city={Solaize},
            postcode={69360} , 
            country={France}}
            
\author[inst1,inst3]{M. C. Righi}
\ead{clelia.righi@unibo.it}
\affiliation[inst3]{organization={Department Mechanical Engineering, Imperial College London},
            addressline={South Kensington Campus}, 
            city={London},
            postcode={SW7 2AZ}, 
            country={UK}}

\begin{abstract}
In pursuit of sustainable lubricant materials, the \textit{in situ} formation of graphitic material has been shown to effectively reduce friction at metallic interfaces. Aromatic molecules are perfect candidates for the formation of carbon-based tribofilms due to their inertness and chemical structure. We selected a group of common aromatic compounds, which are still unexplored in tribology, and we investigated their capability to reduce the adhesion of sliding iron interfaces. Ab initio molecular dynamics simulations show that hypericin, a component of St. John's wort, can effectively separate the mating iron surfaces better than graphene. The size of this molecule, combined with the reactivity and the hindrance of its moieties, plays an important role in maintaining large interfacial distances. Stacked hypericin molecules can easily slide on top of each other due to the intermolecular repulsion arising in the presence of load. The decomposition of the lateral groups of hypericin observed in the dynamic simulations suggests that the polymerization of several molecules can occur in tribological conditions. All these results pave the way for promising alternatives to commonly employed friction modifiers.
\end{abstract}




\end{frontmatter}


\section{Introduction}

Friction is a common phenomenon that occurs every time two surfaces are in relative motion, causing huge energetic losses and CO$_2$ emissions~\cite{HolmbergErdemir2017}. The technologies available today to limit the impact of friction on the environment are based on lubricant materials, e.g. solid lubricants, that are deposited on the surfaces before sliding~\cite{solidreview}, and lubricant additives, that are introduced in the contact through a liquid medium~\cite{Spikes2015}. In particular, the functionality of boundary lubrication additives resides in their capability to decompose in tribological conditions. The molecular fragments resulting from the decomposition of the additives can recombine into a protective layer called tribofilm, that prevents the adhesive interaction between the sliding parts~\cite{Spikes2015}. The most successful representatives of this group of compounds are zinc dialkyldithiophosphate (ZDDP)~\cite{Spikes2004} and molybdenum dithiocarbamate (MoDTC)~\cite{bec2004}, which showcase outstanding tribological properties at metallic contacts. Despite their wide use, these additives contain sulfur and phosphorus, which represent a threat for the environment because these elements can poison the catalysts used to reduce harmful emissions in car exhausts. Therefore, identifying alternative materials to form effective and sustainable tribofilms is an active field of research.

On the other hand, carbon-based materials are extremely successful in reducing friction and wear in a wide variety of environments~\cite{GRIERSON200712}. Graphene is an ideal material in this context, as revealed by the extremely low coefficients of friction measured not only at the nano-, but also at larger-scale applications~\cite{liu2012,feng2013,BERMAN201431,review19}. Some authors have recently demonstrated the possibility to generate \textit{in situ} graphitic tribofilms. Erdemir et al. showed that molecules constituting the lubricating base oil, as well as methane, can form a tribofilm on top of catalitically-active coatings~\cite{Erdemir2016,ramirez2020}. Zhang et al. observed the formation of amorphous carbon and graphene from ethylene glycol, both experimentally and computationally~\cite{ZHANG2020}. Those findings are promising because they demonstrate the possibility to use new organic molecules in tribology to effectively separate interfaces under mechanical stresses.

Here we consider the properties of several aromatic molecules at the interface of iron to promote the formation of a carbon-based tribofilm, representing the initial steps of the mechanisms that lead to the low friction regimes. Iron was chosen for its technologically relevance as the main constituent of  steel~\cite{loehle2018,fe-modtc}. We took into account hypericin and pseudohypericin, graphene flakes with similar shape to hypericin, quinone and anthraquinone, benzyl benzoate and an organic dye called Oil Blue 35. Hypericin-like molecules are commonly used to treat several illnesses, such as depression and cancer~\cite{CHANG2010,agostinis2002}. Quinones are also employed for medical and pharmaceutical purposes~\cite{malik2016,PATEL2021}, as well as in energy storage applications~\cite{gerhardt2017}. Benzyl benzoate is used as an acaricide~\cite{kalpaklioglu1996}, as a catalyst~\cite{LU2019} and has also been used as a milling agent to produce boron nitride nanosheets~\cite{Deepika2014}. Oil Blue 35, a derivative of anthraquinone, is used as a fuel marker~\cite{desouza2017,tomic2018}. Quinones were considered in this study because emodin, a member of this family of compounds, is a precursor of hypericin, both in natural and artificial synthetic pathways~\cite{zhang2022}. Quinones might play a role as intermediates or side products during the development of graphitic material from hypericin in tribological conditions. Benzyl benzoate is another common compound, comparable in size to the quinones, that could be involved in similar tribochemical reactions. The graphene flakes were chosen to study the effect of the lateral groups of hypericin on the adhesion to the metallic substrate. All these aromatic molecules have not been used in tribological applications yet, and, to the best of our knowledge, their interfacial properties are investigated here for the first time for this purpose.

Our approach is based on static and dynamic density functional theory simulations to take into account the electronic degrees of freedom which are essential for a complete description of this complicated tribochemistry. Our aim is to identify the best performing molecules in separating the iron interface and understand the mechanism through which some of these compounds provide low resistance to sliding.


\section{Methods and Systems}

\subsection{Computational protocol}
\begin{figure}[ht]
\centering
\includegraphics[width=0.4\columnwidth]{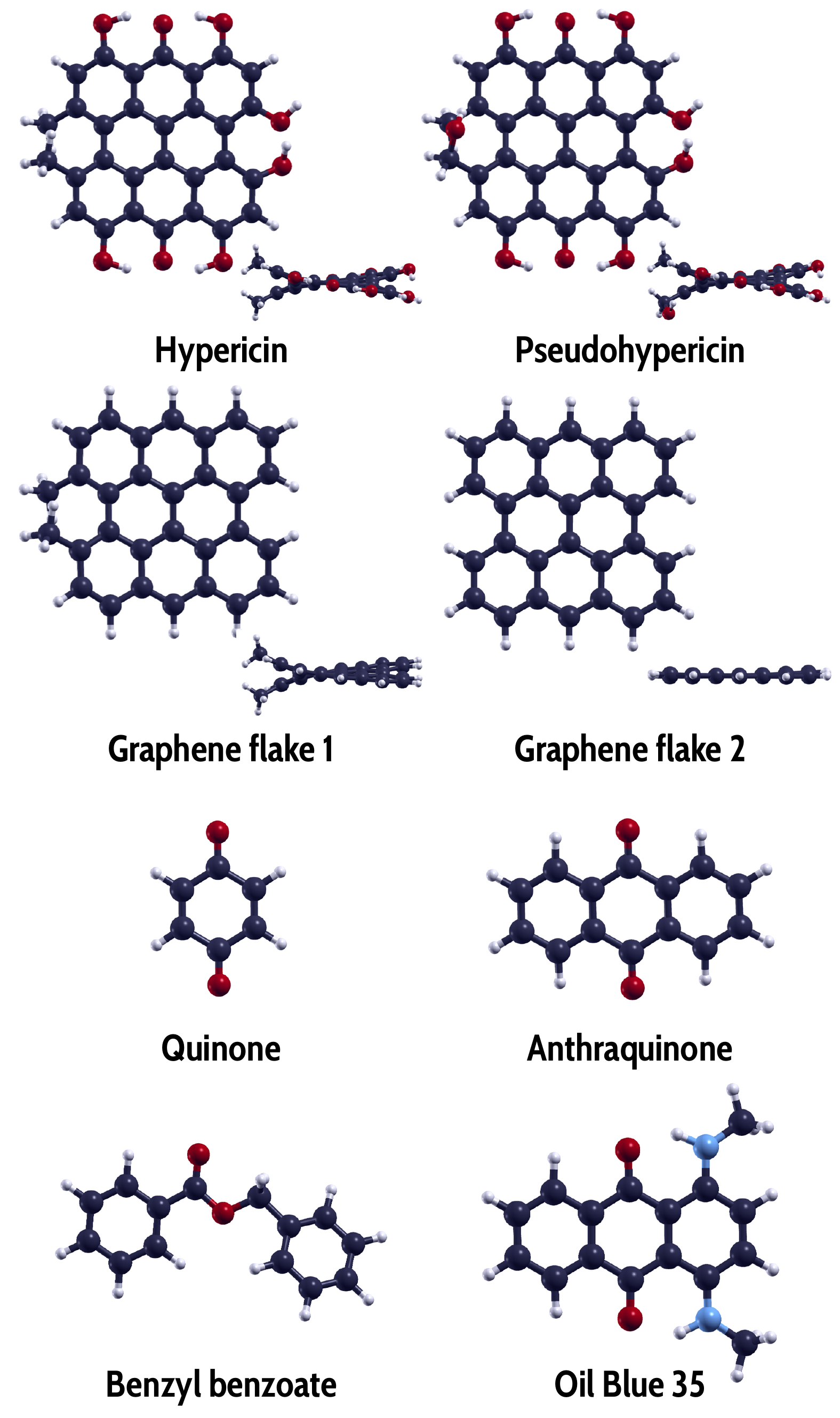}
\caption{Molecular structures of the aromatic compounds considered in this work. Lateral views are also included for the first four molecules, while the latter four are completely flat, similarly to graphene flake 2. White, black, light blue and red atoms correspond to hydrogen, carbon, nitrogen and oxygen, respectively.}\label{fig:1-molecules}
\end{figure}

The chemical structures of the molecules considered in this work are shown in Figure~\ref{fig:1-molecules}. Hypericin is an organic compound composed of eight aromatic rings. Hydroxyl, oxide and methyl groups occupy the edges of the molecule, where several intramolecular hydrogen bonds are present~\cite{ULICNY2000}. Pseudohypericin differs from hypericin for the presence of an additional hydroxyl group attached to a methylene bridge. We considered two graphene flakes composed of eight aromatic rings, as in the case of hypericin. Graphene flake 1 presents two methyl groups in the same position of hypericin, unlike graphene flake 2. The rest of the edges of the graphene flakes are saturated by hydrogen atoms. Finally, we considered benzyl benzoate, quinone, anthraquinone and Oil Blue 35, a derivative of anthraquinone with two alkylamino chains attached to one of the aromatic rings. Typically, these chains contain four carbon atoms~\cite{tomic2018}, yet we considered methylamino groups to reduce the computational effort. We believe this approximation can be justified by the practically identical fragmentation energies of the different alkylamino chains, as shown in the Appendix.

To investigate the properties of these aromatic compounds at the iron interface, we performed DFT simulations using Quantum ESPRESSO~\cite{QE1}. The \emph{ab initio} molecular dynamics simulations were performed using a version of the program modified by our group and previously used to study the tribochemistry of lubricant additives at iron interfaces~\cite{loehle2018}. The Perdew-Burke-Ernzerhof (PBE) exchange-correlation functional~\cite{PBE} was used in combination with Ultrasoft pseudopotentials. Values of 30 and 240 Ry were used as cutoff for the kinetic energy of the plane waves and for the charge density, respectively, in agreement with our previous studies~\cite{fe-modtc,passmodtc}. In all the static calculations, the integration of the charge density was performed on K-point grids that were properly rescaled with respect to the iron bulk, originally sampled with an $ 8\times 8 \times 8$ $\Gamma$-centered grid~\cite{giulio}, to maintain an equivalent K-point density of the Brillouin zone. Spin polarization was taken into account, and a Gaussian smearing of 0.02 Ry was added to better describe the electronic occupations around the Fermi level.

\begin{figure}[ht]
\centering
\includegraphics[width=\columnwidth]{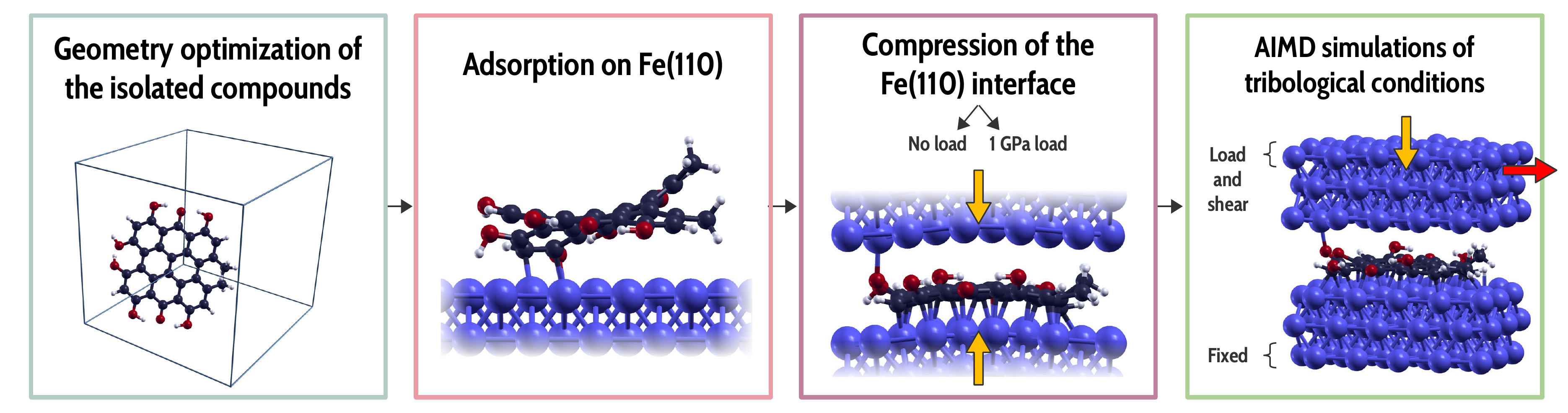}
\caption{Computational protocol adopted in this work.}\label{fig:2-protocol}
\end{figure}

We followed the computational protocol summarized in Figure~\ref{fig:2-protocol}.

The initial molecular geometries were generated by using the Avogadro software~\cite{avogadro1}. The molecules were then optimized in a cubic cell with an edge of 40 angstrom to avoid interactions between periodic replicas. All the geometry optimizations in this work were performed using the BFGS algorithm~\cite{bfgs} and they were stopped when the total energy and the forces converged under the threshold of $1\times10^{-4}$ Ry and $1\times10^{-3}$ Ry/Bohr, respectively.

For molecular adsorption, Fe(110) slabs with different sizes were considered. The size of the slab used to adsorb the molecules was $5\times3\sqrt{2}$, with a thickness of three atomic layers and the coordinates of the bottom layer fixed, except for anthraquinone and quinone, for which four-layers thick $4\times3\sqrt{2}$ and $3\times3\sqrt{2}$ iron slabs were used, respectively, without any geometric constraint. A vacuum region of at least 10 angstrom was included in the system as a compromise between minimizing the interaction among periodic replicas in the vertical direction and keeping the computational effort feasible.

Since dispersion forces may play an important role in the systems considered in this work, we included the DFT-D3 correction~\cite{grimmed3} in all the calculations. To estimate the strength of the interactions arising from dispersion forces, we also calculated the adsorption energies without dispersion correction. The adsorption energies were calculated as:

\begin{equation}\label{eqn:adsorption_energy}
  E_{ads} = E_{tot} - E_{mol} - E_{iron}
\end{equation}
\noindent
where $E_{tot}$, $E_{mol}$ and $E_{iron}$ are the total energies of the whole adsorbed system, of the isolated molecule, and of the isolated iron slab, respectively.

Variable-cell calculations were performed, starting from the geometries obtained from the adsorption simulations, to prepare the systems for the subsequent dynamic simulations. In the variable-cell calculations, we first optimized the geometry of the system without any external forces. In this way, the molecules were confined at an iron interface. Afterwards, a compressive stress of 1 GPa was applied to the system in the vertical direction, while keeping the lateral dimensions constant. Both a single and a double layer of intercalated molecules were considered, in analogy to a previous study concerning graphene enclosed between iron surfaces~\cite{restuccia2016}. Studying the behavior of stacked molecules is also useful to represent the aggregation phenomenon typically observed in hypericin~\cite{bano2011}. To compare the surface coverage of the compounds with different sizes, we normalized the number of atoms in each molecule by the number of atoms composing the surface, as follows:

\begin{equation}\label{eqn:coverage}
  \theta = \frac{n_C + n_N + n_O}{n_{Fe}}
\end{equation}
\noindent
where $n_C$, $n_N$ and $n_O$ are the number of C, N and O atoms in the molecule and $n_{Fe}$ is the number of Fe surface atoms. Calculating an effective size of a molecule is not trivial due to the complicated shape of its potential energy surface. Equation~\ref{eqn:coverage} is a simplified way to compare the surface areas occupied by the different aromatic compounds. We will refer to $\theta$ as atomic coverage in the following. In the case of benzyl benzoate, we included two adjacent molecules in the same simulation cell to reach an atomic coverage comparable to the one of hypericin. Additionally, we considered one and two molecules of quinone per layer to investigate the role of the coverage in the separation of the interface.

One can estimate the interfacial adhesion as follows:

\begin{equation}\label{eqn:int_adhesion_energy}
  E_{adh} = E_{contact} - E_{separated}
\end{equation}
\noindent
where $E_{contact}$ and $E_{separated}$ correspond to the total energies of the compressed system resulting from the variable-cell calculation and the same system in which the two periodic replicas are separated by at least 10 angstrom of vacuum. The average interfacial distance was also calculated at the end of the variable-cell calculations.

Finally, Born-Oppenheimer molecular dynamics simulations were performed, taking the compressed systems of hypericin, the graphene flakes and quinone from the previous step as initial geometries. In these simulations, the interface was composed of two iron tri-layers, except in the case of quinone, for which four iron layers were considered in each slab. The integration of the charge density was carried out at the $\Gamma$-point, and a time step of approximately 1.45 fs was used. The first picosecond of the dynamic simulations was spent on the equilibration of the systems at two different temperatures, namely 300 and 380 K, using the velocity rescaling algorithm. These temperatures were chosen because they represent a typical range of operation in experiments focusing on the tribochemistry of organic compounds~\cite{ramirez2020,Long2022}. Vertical forces corresponding to a normal load of 1 GPa were applied to the topmost iron layer since the beginning of the simulation. After the equilibration time, a lateral velocity of 1 \AA/ps in the [1$\overline1$0] direction was applied to all the atoms of the topmost iron layer. The coordinates of the bottom iron layer were kept fixed throughout the whole simulation.

For each dynamic simulation, the average interfacial distance $z_{eq}$, the average interaction energy at the interface and the average resistive force were calculated. The average interaction energy can be obtained in the following way:

\begin{equation}\label{eqn:int_adhesion_energy_dyn}
  \langle E_{int}(t) \rangle = \langle E_{tot}(t) \rangle - E_{top} - E_{bottom}
\end{equation}
\noindent
where $E_\mathrm{tot}$, $E_\mathrm{top}$ and $E_\mathrm{bottom}$ correspond to the total energy of the system at each step of the simulation and the total energies of the optimized top and bottom parts of the interface calculated separately, respectively. The average resistive force per unit area was calculated as:

\begin{equation}\label{eqn:lat_forces_dyn}
  \langle F_{lat}(t) \rangle = \frac{1}{A} \sum_i \langle {f_x^i(t)} \rangle
\end{equation}
\noindent
where $A$ is the surface area of the simulation cell and $f_x^i(t)$ is the $x$-component of the force of the $i$-th atom, being $x$ the direction of the sliding motion.


\section{Results and Discussion}

\subsection{Adsorption on Fe(110) and compression of the interface}

\begin{figure}[ht]
\centering
\includegraphics[width=0.5\columnwidth]{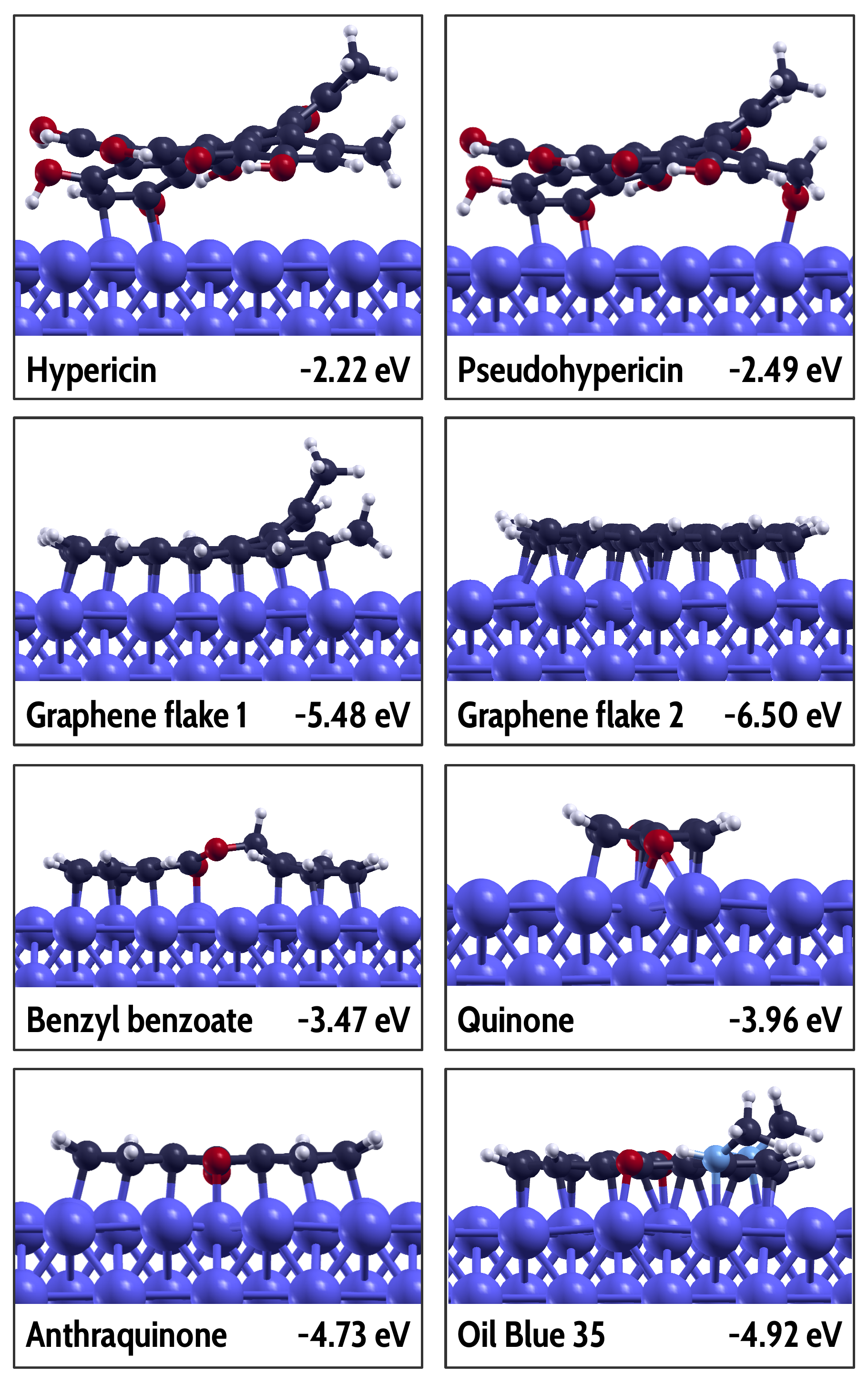}
\caption{Lateral view of the adsorbed molecules on Fe(110). The corresponding adsorption energy in eV is reported below each system.}\label{fig:3-adsorption}
\end{figure}

All the molecules considered in this work strongly attach to the Fe(110) surface. The geometries and the adsorption energies of the different compounds are shown in Figure~\ref{fig:3-adsorption}. Despite being the largest molecules, hypericin and pseudohypericin are characterized by the lowest adsorption energies, due to their twisted molecular structures. The other molecules are flat, allowing for larger interactions with the substrate. Graphene flake 2 has the strongest adsorption energy, being -6.5 eV. By normalizing the adsorption energies by the number of atoms in the molecules, quinone turns out to be the most strongly attached, with an adsorption energy of -0.33 eV/atom. The dispersion forces stabilize the adsorption of these compounds with contributions in the range 31--70 meV/atom, which represent the cases of hypericin and anthraquinone, respectively.

\begin{figure}[ht]
\centering
\includegraphics[width=0.9\columnwidth]{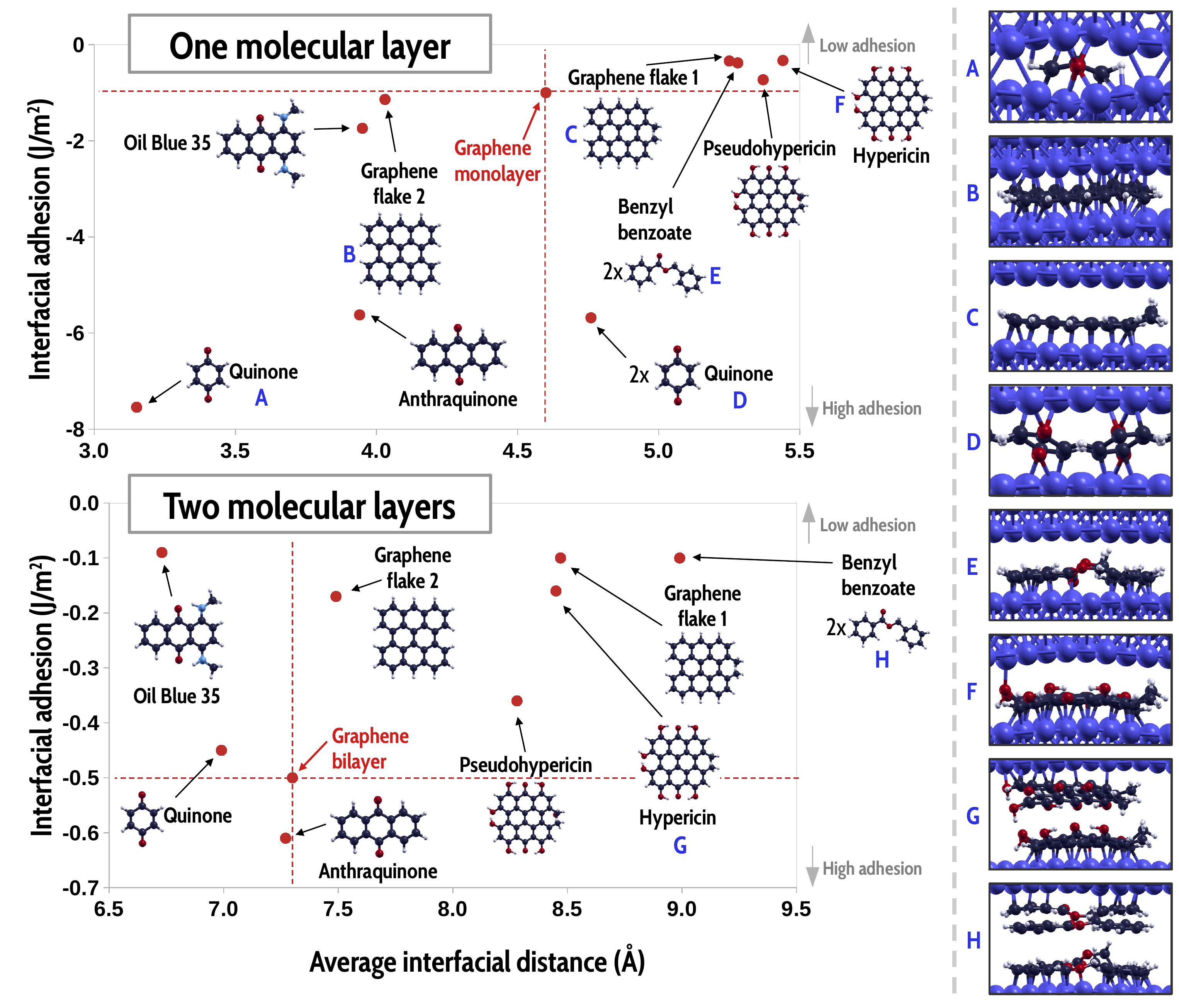}
\caption{Interfacial adhesion of the compressed systems with respect to their interfacial distance, in the absence of load. The top and bottom plots correspond to a single and two stacked molecular layers, respectively. Each dot in the plots is connected to the respective compound and the red dashed lines highlight the points relative to a graphene monolayer and bilayer. Lateral views of selected systems are included on the right side of the figure.}\label{fig:4-confinement}
\end{figure}

The final geometries of the adsorbed systems were used as the starting point of the variable-cell calculations in which the molecules are enclosed at the Fe(110) interface. The interfaces were simulated both in the absence of load, in analogy with a previous study concerning the lubricating properties of graphene mono- and bilayers~\cite{restuccia2016}, and in the presence of a load of 1 GPa. Since the molecules are characterized by different sizes, the interfacial coverage must be taken into account to compare the results of these calculations. In the case of benzyl benzoate and quinone, two molecules are necessary to achieve an atomic coverage comparable to the one provided by the larger compounds. Additionally, we compared the capability of a single and a double molecular layer to separate the interface, with the second layer stacked on top of the first one. Table~\ref{tab:1-confinement} reports the interfacial distances and adhesion energies for one and two molecular layers in the absence of load. The corresponding values at 1 GPa are included in the Appendix. The interfacial distances at 1 GPa are slightly smaller than the ones in the absence of load, while the interfacial adhesion is generally stronger. Figure~\ref{fig:4-confinement} shows selected geometries of the optimized interfaces without load and the plots of the interfacial adhesion with respect to the interfacial distance for all the systems.

\begin{table}[ht]
\small
\caption{Atomic coverage $\theta$, interfacial distance $z_{eq}$ and interfacial adhesion $E_{adh}$ for one and two molecular layers at the Fe(110) interface optimized without load. The atomic coverage is defined in Equation~\ref{eqn:coverage} as the ratio between the atoms of the molecules heavier than hydrogen and the number of iron atoms composing the slab.}
\label{tab:1-confinement}
\begin{center}
\begin{tabular*}{\textwidth}{@{\extracolsep{\fill}}lccccc}
\hline
\multirow{2}{*}{} & \multirow{2}{*}{} & \multicolumn{2}{c}{\rule{0pt}{10pt} One intercalated layer} & \multicolumn{2}{c}{Two intercalated layers} \\
\cmidrule(lr){3-4} \cmidrule(lr){5-6}
Molecule & $\theta$ & $z_{eq}$ (\AA) & $E_{adh}$ (J/m$^2$) & $z_{eq}$  (\AA) & $E_{adh}$ (J/m$^2$) \\
\midrule
Hypericin          & 126.6\% & 5.44 & -0.33 & 8.45 & -0.16 \\
Pseudohypericin    & 130.0\% & 5.37 & -0.73 & 8.28 & -0.36 \\
Graphene flake 1   & 100.0\% & 5.25 & -0.34 & 8.47 & -0.10 \\
Graphene flake 2   & 93.3\%  & 4.03 & -1.14 & 7.49 & -0.17 \\
2x Benzyl benzoate & 106.6\% & 5.28 & -0.38 & 8.99 & -0.10 \\
Quinone            & 66.6\%  & 3.15 & -7.54 & 6.99 & -0.45 \\
2x Quinone         & 133.3\% & 4.76 & -5.68 & 7.07 & -0.87 \\
Anthraquinone      & 100.0\% & 3.94 & -5.62 & 7.27 & -0.61 \\
Oil Blue 35        & 66.6\%  & 3.95 & -1.74 & 6.73 & -0.09 \\
\hline
  \end{tabular*}
  \end{center}
\end{table}

Different factors determine the interfacial properties of these systems. The strongest influence on the interfacial distance and adhesion is determined by the presence of either one or two stacked molecules at the interface. Indeed, the interfacial distance increases by 3.5 \AA \,on average, when two molecules are stacked at the interface. As shown in Figure~\ref{fig:4-confinement}, two hypericin molecules, as well as two layers of benzyl benzoate, interact strongly with the surfaces and weakly with each other. This behavior is observed consistently across all the systems, and is useful for reducing the resistance to sliding.

The conformation of the molecules also plays an important role, as demonstrated by the graphene flakes. The interfacial separation generated by graphene flake 1, which has two methyl groups in the same position as hypericin, is 1.2 \AA \,wider than the one of graphene flake 2, in which the methyl groups are absent. The methyl groups at the edge of the graphene flake repel the countersurface due to their geometry, dictated by the $sp^3$ hybridization of the carbon atom. The interfacial adhesion is 0.8 J/m$^2$ higher in the case of graphene flake 2, indicating that graphene flake 1 offers less resistance to slide the Fe(110) interface.

The interfacial coverage of the molecules also influences separation and adhesion. In the case of quinone, which is a small and flat molecule compared to the other ones considered in this study, several bonds are formed across the interface. Even though the atomic coverage of quinone and Oil Blue 35 is similar, the size of the latter molecule prevents the formation of interfacial bonds. Indeed, considering two adjacent quinones at the interface prevents the formation of Fe-Fe bonds, despite the strong interactions between the molecules and the iron surfaces that keep the interfacial adhesion relatively high.

The results for all the molecules are summarized in the plots of the interfacial adhesion with respect to interfacial distance, shown in Figure~\ref{fig:4-confinement}. The values of the graphene mono- and bilayer are also included as a reference for the cases of a single and double molecular layer, respectively~\cite{restuccia2016}. The molecules with ideal properties are found in the top-right region of the plots, as they provide weaker interfacial adhesion and larger interfacial separations compared to graphene. Hypericin outperforms all the other compounds in the case of a single molecular layer at the interface. The interfaces separated by two molecular layers of hypericin, benzyl benzoate, Oil Blue 35 and the graphene flakes are all characterized by very low adhesion.

\subsection{Dynamics at the Fe(110) interface}

\begin{figure}[ht]
\centering
\includegraphics[width=0.8\columnwidth]{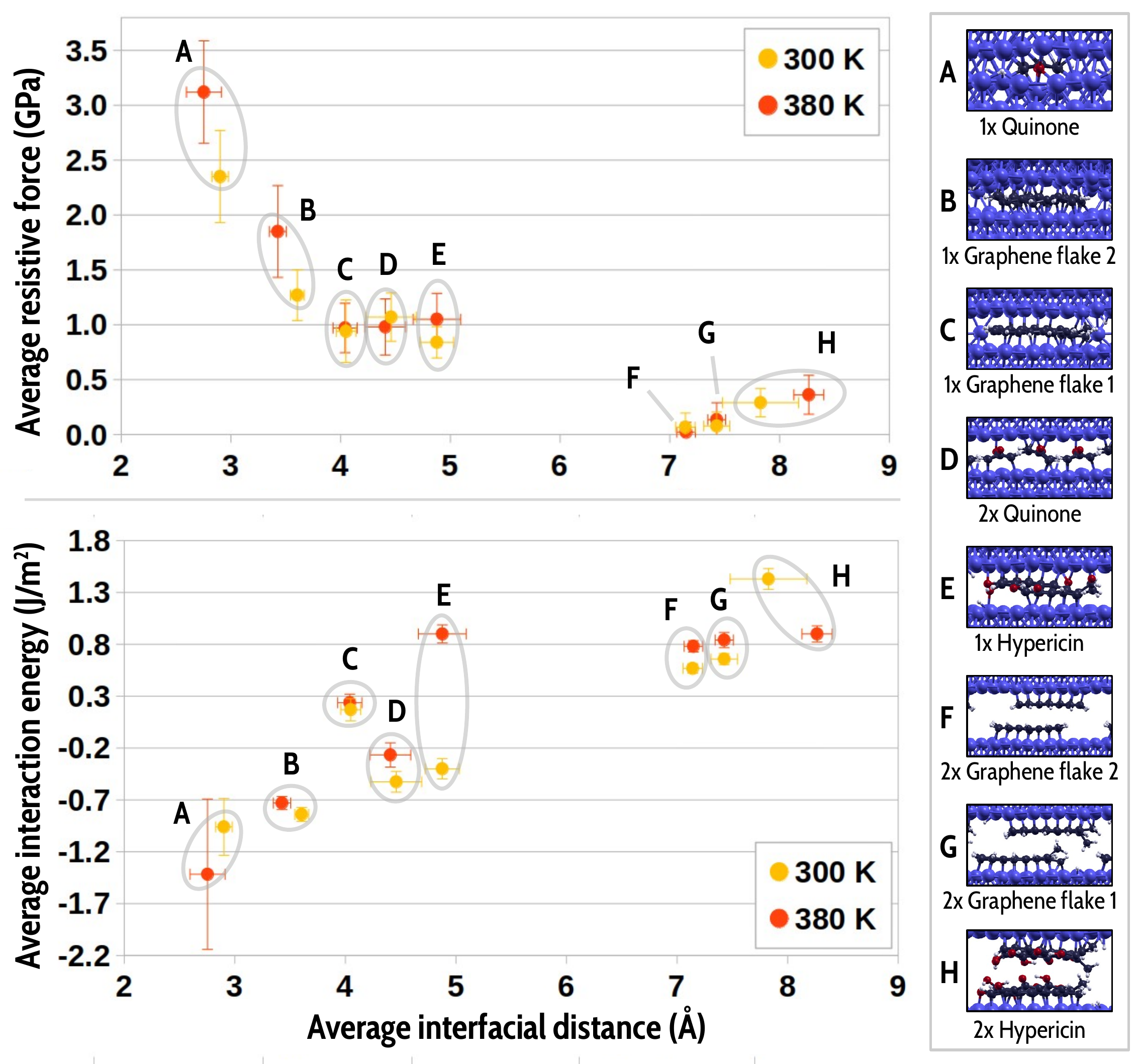}
\caption{Resistive force and interfacial interaction energy with respect to the interfacial distance averaged over the dynamic simulations. The insets on the right show the lateral views of the interfaces after 4 ps. The error bars on the resistive force were calculated using the standard deviation from a block analysis, as in Ref.~\citenum{templeton2021}.}\label{fig:5-graphs-dynamics}
\end{figure}

Figure~\ref{fig:5-graphs-dynamics} shows the resistive force and the interfacial interaction energy, averaged over the dynamic simulation after the equilibration time, with respect to the average interfacial distance. The systems with a single layer of molecules at the interface are characterized by high resistance to sliding, strong adhesive forces and low interfacial separations. We summarize the main events observed during the dynamic simulations in the following, starting from the systems with the lowest interfacial separation.

\begin{figure}[ht]
\centering
\includegraphics[width=\columnwidth]{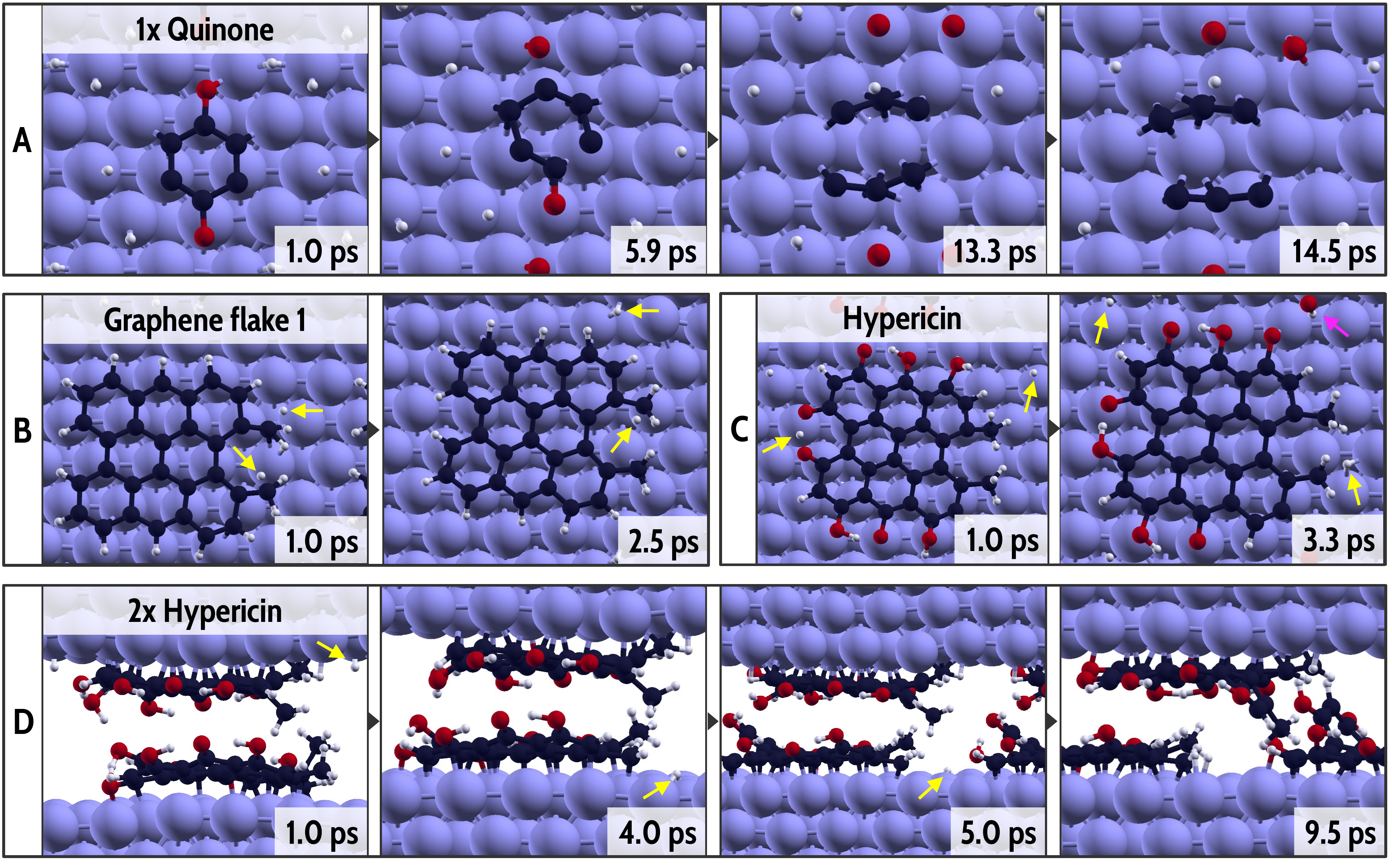}
\caption{Snapshots of the dynamic simulations of selected systems at 300 K: (A) top views of the decomposition of quinone; (B) and (C) top views of graphene flake 1 and hypericin, respectively; (D) lateral views of the sliding of two hypericin molecules. The size of the iron atoms in all the panels was increased by approximately 40\%, while their color was altered to improve the visibility of the molecular fragments. The yellow and purple arrows in panels b-d indicate dissociated H atoms and hydroxyl groups, respectively.}\label{fig:6-final-dynamics}
\end{figure}

Hydrogen atoms usually detached in the first picoseconds of the simulations. In the case of a single quinone at 300 K, all the C-H bonds broke before 1 ps. The first C-O bond broke at approximately 5.9 ps, and full fragmentation of the molecule occurred at 13.3 ps, as shown in Figure~\ref{fig:6-final-dynamics}A. At 380 K, these dissociation occurred faster. The first C-O bond breaking and the full fragmentation occurred at 3.7 and 6.2 ps, respectively. In both simulations several Fe-Fe bonds across the interface were formed, leading to the highest resistive forces among all the systems. The total simulation time of this system is 15 ps for both temperatures, and representative snapshots of the system at 380 K are reported in the Appendix.

Graphene flake 2 did not effectively separate the iron interface, causing the formation of many interfacial bonds. The dissociation of the C-H bonds, occurred in the first 2 ps, was the only event observed during the simulations at 300 and 380 K, which lasted 4.1 and 4.5 ps, respectively. Representative snapshots of graphene flake 2 at the sliding interfaces are shown in the Appendix. Graphene flake 1 was able to better separate the interface thanks to the additional methyl groups on the side. Indeed, the average interaction energy of graphene flake 1 is negative, i.e. repulsive, while the corresponding value for graphene flake 2 is negative, i.e. attractive. These methyl groups were pushed down by the top surface, which caused the decomposition of one H atom per group, as shown in Figure~\ref{fig:6-final-dynamics}B, and the rehybridization of carbon from $sp^3$ to $sp^2$ during the first 2 ps of the simulations. Less Fe-Fe bonds were formed compared to graphene flake 2 due to the increased surface area of the flat graphene flake 1. The simulations for graphene flake 1 lasted 8.6 and 10.6 ps.

As expected from the static calculations, putting two adjacent quinone molecules provided a larger separation of the interface. Both at 300 and 380 K, the two molecules remained undecomposed and no Fe-Fe bonds were formed during the simulations, as shown in the Appendix. The simulations of this system lasted 15 ps.

Among the simulations containing a single intercalated layer of molecules, hypericin showcased the most promising performance. The interfacial distance was the highest compared to the previous systems, the resistive force was comparable to the one of graphene flake 1 and two quinone molecules, while the interaction at the interface was positive at 380 K, indicating that the compressed system was less favorable than the adsorbed molecule on iron and the top surface at infinite distance. At 300 K, hydrogen atoms from lateral hydroxyl and methyl groups detached at approximately 0.9 and 3.3 ps, respectively. At 1.2 ps, one of the oxygen atoms composing an hydroxyl group detached from hypericin and diffused on the top surface. These decompositions are visible in Figure~\ref{fig:6-final-dynamics}C. At 380 K, an hydroxyl group detached before 1 ps and other hydrogen atoms were removed from hypericin after 7.4 ps, as shown in the Appendix. The simulations for a single layer of hypericin molecules lasted 10.3 and 9.8 ps in total.

Stacked graphene flakes 1 and 2 and hypericin molecules provided high interfacial separation, low resistive forces and positive interaction energies, with the double hypericin layer outperforming the other two systems. The stacked graphene flakes remained completely intact during the whole simulations. The two hypericin molecules did not chemically react either, yet in the simulation at 300 K the lateral groups of the molecules got stuck at 4.6 ps, causing a bending of the two structures and an increase in the interfacial repulsion with respect to the same system at 380 K (Figure~\ref{fig:6-final-dynamics}D). The same bending was be observed only after 7 ps at 380 K. The simulations for all these systems lasted approximately 10 ps on average, and representative snapshots are also included in the Appendix.

To summarize, the molecules considered in this work separate the iron interface differently. The following trends can be observed in the plots in Figure~\ref{fig:5-graphs-dynamics}. At low interfacial distances, the resistive force is high and the average interaction energy is negative, indicating that the compressed system is favorable, while the reverse is true at high interfacial distances. The best performing molecule at the interface of iron is hypericin, both in the case of a single and a double intercalated layer, as it provided the highest interfacial separation and the least resistance to sliding. The effect of temperature did not clearly emerge from the dynamic simulations. The interfaces at 380 K generally showcased slightly higher resistive forces and more repulsive interactions than the ones at 300 K, yet the main effect of the temperature on the simulations is simply to accelerate the chemical decomposition of the compounds. Instead, the steric hindrance of the molecules have a stronger influence on the interfacial properties. Flat molecules cannot effectively separate the sliding surfaces, and low coverage values can induce the formation of interfacial Fe-Fe bonds. The lateral groups of these compounds play an important role also for the following steps of the reaction that may eventually lead to the formation of a graphitic tribofilm.


\section{Conclusion}

The properties of several aromatic molecules at the Fe(110) surface and interface were evaluated for the first time with ab initio molecular dynamics simulations, with a particular interest in their capability to separate sliding interfaces. Our simulations highlighted the role of hypericin as a possible sustainable lubricant, as it effectively reduces the adhesive friction of the iron interface. We found that an efficient passivation of the sliding surfaces can be obtained by stacked undissociated molecules, which strongly repel each other. The comparison with other aromatic compounds reveals that these beneficial interfacial properties stem from the size of the molecule and the steric hindrance of its lateral groups. Indeed, a layer of these molecules can efficiently reduce adhesive friction of the Fe(110) surface even better than graphene.

The effect of temperature was also taken into account in the dynamic simulations. At 380 K, the dissociation of the lateral groups of hypericin is generally accelerated with respect to the corresponding system at 300 K, suggesting that larger graphenic patches may be formed by the coalescence of different molecules exposing dangling bonds.

Even though we only considered the most stable surface of iron in this study, we believe that the lubrication mechanisms of these compounds could be valid for other surface orientations, onto which the aromatic molecules can adsorb more strongly. On the other hand, the presence of surface oxides can limit the adsorption energy of the compounds, yet the slipperiness of undissociated hypericin molecules and their potential to build larger carbon structures are a promising perspective that deserves further investigation in future works.
    

\section{Acknowledgments}
The authors thank Benoit Thiebaut, Yun Long, Maria Isabel De Barros Bouchet and Jean Michel Martin for fruitful discussions. The European Research Council (ERC) under the European Union's Horizon 2020 Research and Innovation Programme (Grant Agreement No. 865633) is acknowledged. The geometries of the systems in this work were represented using XCrySDen~\cite{xcrysden}.

\appendix

\section{Fragmentation energies of hypericin, benzyl benzoate, Oil Blue 35}

\begin{figure}[ht]
\centering
\includegraphics[width=0.5\columnwidth]{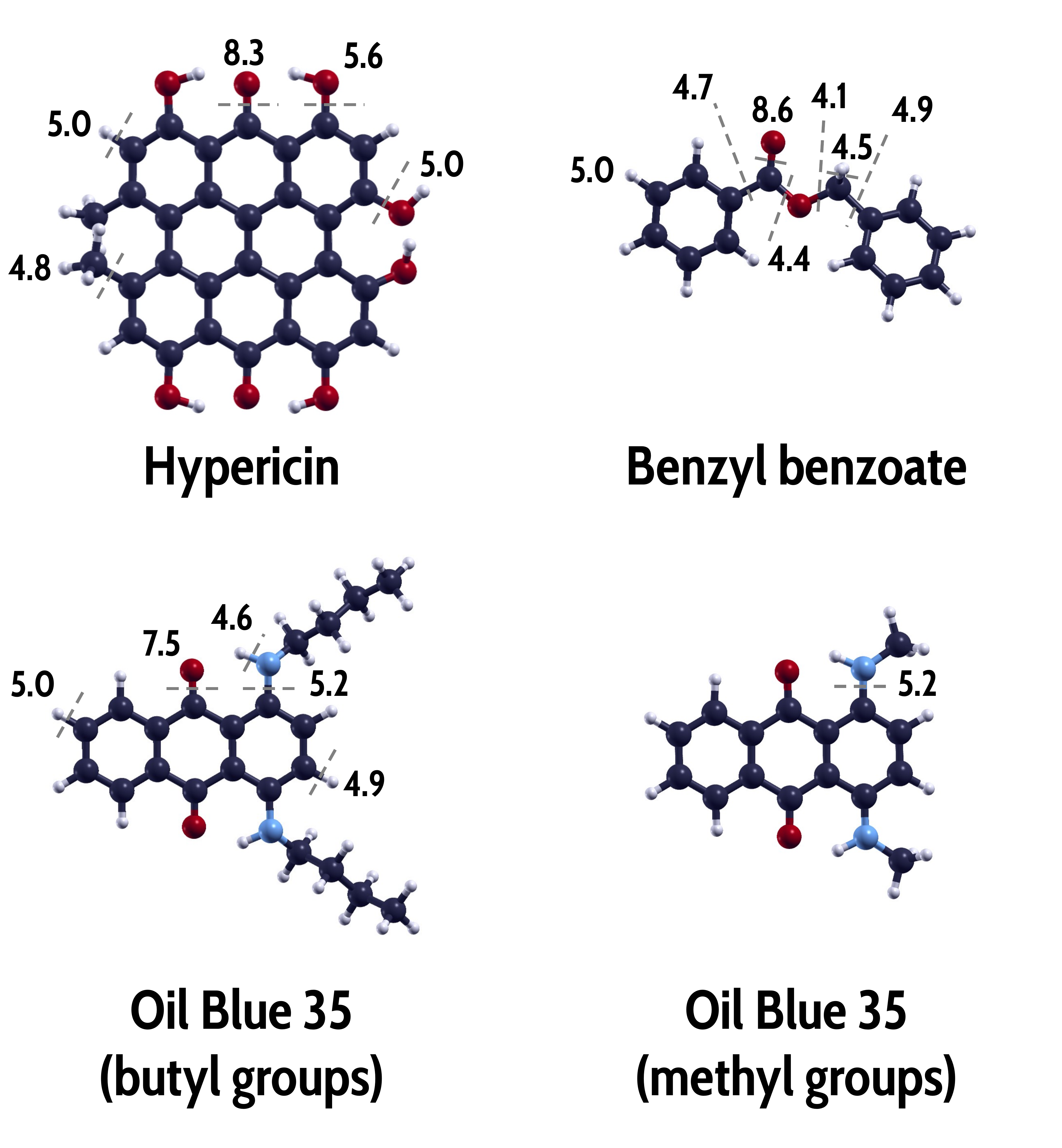}
\caption{Fragmentation energies, calculated as in Eqn.~\ref{eqn:s1}, of hypericin, benzyl benzoate and Oil Blue 35. All the energy values are expressed in eV.}\label{fig:s1}
\end{figure}

The fragmentation energies shown in Figure~\ref{fig:s1} were calculated as:

\begin{equation}\label{eqn:s1}
  \Delta E_\mathrm{frag} = E_\mathrm{frag1} + E_\mathrm{frag2} - E_\mathrm{molecule}
\end{equation}
\noindent
where $E_\mathrm{frag1}$, $E_\mathrm{frag2}$ and $E_\mathrm{molecule}$ are the total energies of the two molecular fragments and the whole molecule, respectively, in analogy with our previous studies~\cite{static-modtc,niacac2,mono-modtc}.

\section{Details on the compressed molecules at the Fe(110) interface}

\begin{figure}[ht]
\centering
\includegraphics[width=\columnwidth]{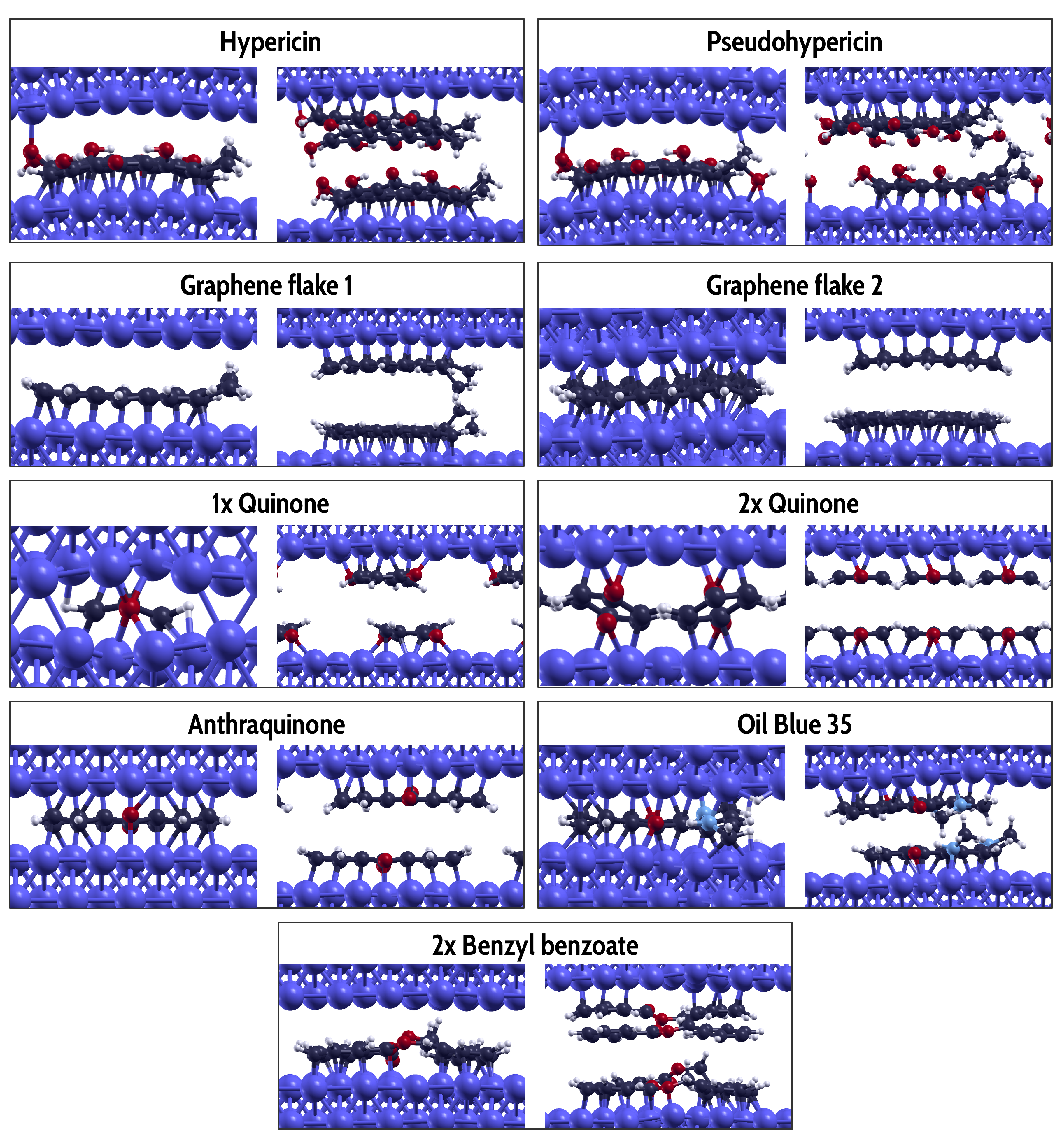}
\caption{Lateral views of the interfaces with intercalated molecules in the absence of load. For each system, the interfaces with one and two molecular layers are shown.}\label{fig:s2}
\end{figure}

Figure~\ref{fig:s2} includes the lateral views of all the systems without load.

On average, the interfacial distances at 1 GPa are 0.26 and 0.51 \AA\, smaller than the corresponding values without load in the case of one and two molecular layers, respectively. The interfacial adhesion at 1 GPa for one layer of molecules is more favorable by 0.14 J/m$^2$ with respect to the uncompressed system, except for two quinone molecules, for which the interfacial adhesion becomes less favorable by 0.85 J/m$^2$. For two molecular layers, the interfacial adhesion at 1 GPa does not significantly change with respect to the uncompressed systems, except for quinones, for which the interaction becomes more favorable by 1.8 eV J/m$^2$.

\begin{table}[ht]
\small
\caption{Atomic coverage $\theta$, interfacial distance $z_{eq}$ and interfacial adhesion $E_{adh}$ for one and two molecular layers at the Fe(110) interface, optimized at 1 GPa.}
\label{tab:2-load}
\begin{center}
\begin{tabular*}{\textwidth}{@{\extracolsep{\fill}}lccccc}
\hline
\multirow{2}{*}{} & \multirow{2}{*}{} & \multicolumn{2}{c}{\rule{0pt}{10pt} One intercalated layer} & \multicolumn{2}{c}{Two intercalated layers} \\
\cmidrule(lr){3-4} \cmidrule(lr){5-6}
Molecule & $\theta$ & $z_{eq}$ (\AA) & $E_{adh}$ (J/m$^2$) & $z_{eq}$  (\AA) & $E_{adh}$ (J/m$^2$) \\
\midrule
Hypericin          & 126.6\% & 5.27 & -0.40 & 7.90 & -0.19 \\
Pseudohypericin    & 130.0\% & 5.16 & -0.79 & 8.04 & -0.16 \\
Graphene flake 1   & 100.0\% & 5.06 & -0.34 & 7.63 & -0.15 \\
Graphene flake 2   & 93.3\%  & 3.91 & -1.35 & 7.29 & -0.16 \\
2x Benzyl benzoate & 106.6\% & 4.30 & -0.57 & 7.76 & -0.15 \\
Quinone            & 66.6\%  & 3.11 & -7.71 & 6.78 & -2.25 \\
2x Quinone         & 133.3\% & 4.37 & -4.83 & 7.04 & -0.69 \\
Anthraquinone      & 100.0\% & 3.87 & -5.69 & 7.08 & -0.59 \\
Oil Blue 35        & 66.6\%  & 3.78 & -1.93 & 7.02 & -0.76 \\
\hline
  \end{tabular*}
  \end{center}
\end{table}

\clearpage

\section{Additional snapshots of the dynamic simulations}

\begin{figure}[ht]
\centering
\includegraphics[width=\columnwidth]{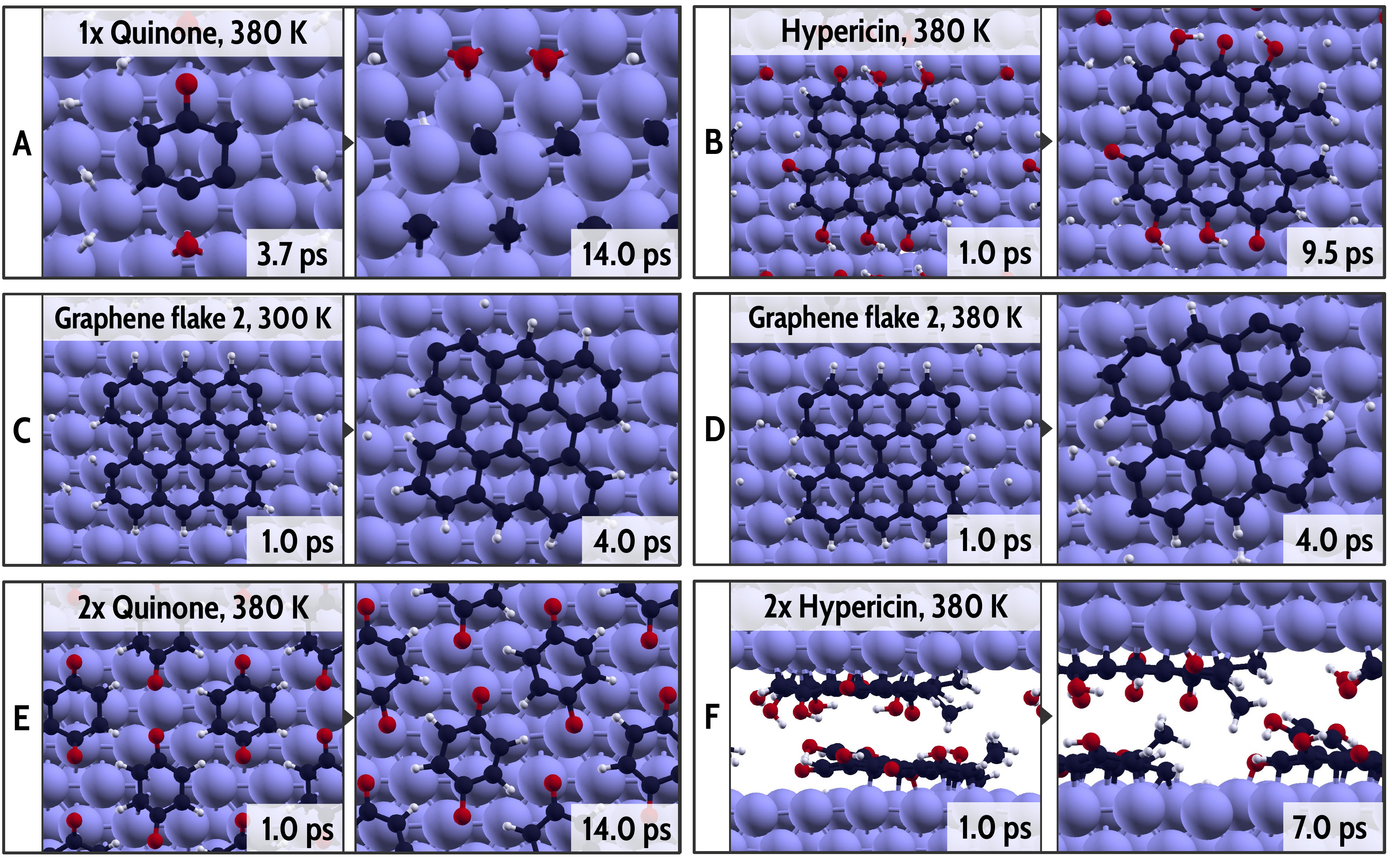}
\caption{Snapshots of the dynamic simulations of selected systems: (A) top views of the decomposition of quinone at 380 K; (B) top views of hypericin at 380 K; (C) and (D) top views of graphene flake 2 at 300 and 380 K, respectively; (E) top views of two adjacent quinone molecules at 380 K; (F) top and lateral views of two stacked hypericin molecules. The size of the iron atoms in all the panels was increased by approximately 40\%, while their color was altered to improve the visibility of the molecular fragments.}\label{fig:s3}
\end{figure}

\bibliographystyle{elsarticle-num-names} 
\bibliography{bibliography}

\end{document}